\documentclass[12pt]{article}
\usepackage{epsf}
\begin{document}
\begin{titlepage}
\title{Koide's $Z_3$-symmetric parametrization, quark masses and mixings}
\author{
{Piotr \.Zenczykowski }\footnote{E-mail: piotr.zenczykowski@ifj.edu.pl}\\
{\em Division of Theoretical Physics}\\
{\em The Henryk Niewodnicza\'nski Institute of Nuclear Physics}\\
{\em Polish Academy of Sciences}\\
{\em Radzikowskiego 152,
31-342 Krak\'ow, Poland}\\
}
\maketitle
\begin{abstract}
The sets of charged-lepton ($L$) and quark ($D,U$) masses may be param\-etrized in a 
$Z_3$-symmetric language appropriate for the discussion of Koide's formula. 
Experiment suggests that at the low-energy scale the relevant phase parameters
$\delta_f$ take on possibly exact values of
 $\delta_L=3\delta_D/2=3\delta_U=2/9$. 
For $k_f$ (the other parameter relevant for the pattern of masses), 
 a similarly simple expression ($k_L=1$)
is known for charged leptons only. 
 Using the Fritzsch-Xing
 decomposition of quark-mixing matrices, we show that the suggested
 pattern of low-energy quark masses is consistent 
with an earlier conjecture that $k_{D,U} \approx 1 $ in the weak basis.

\end{abstract}

\vfill
\end{titlepage}

\section{The doubly special Koide's parametrization}
The appearance of three generations of leptons and quarks
and the related issue of their masses started to baffle us three 
quarters of a century ago. Over the years the problem has become further 
complicated by the presence of inter-generation mixing, as revealed
 in weak interactions. Fortunately, 
 various approximate regularities have been found
in the observed pattern of particle masses and mixings. 
Among the many possible parametrizations of these regularities, there
might be some whose simplicity could help us in deciphering  
physics beyond the Standard Model. \\

One of the most interesting of such regularities is an empirical relation
between the charged-lepton masses discovered by Koide
\cite{Koide} (for a brief review see \cite{RiveroGsponer}): 
\begin{equation}
\label{KoideFormula}
\frac{m_e+m_{\mu}+m_{\tau}}{(\sqrt{m_e}+\sqrt{m_{\mu}}+\sqrt{m_{\tau}})^2}=
\frac{1+k_L^2}{3},
\end{equation}
with $k_L$ equal exactly $1$. 
When the experimental $e$ and $\mu$ masses (here taken from \cite{PDG})
are inserted into
Eq. (\ref{KoideFormula}), this relation predicts the tauon mass within one standard 
deviation from its observed value:
\begin{eqnarray}
\label{tauKoideprediction}
m_{\tau}({k_L=1})&=&1776.9689~{\rm MeV}\\
m_{\tau}({\rm exp})&=&1776.82\pm 0.16~{\rm MeV}.
\end{eqnarray}
Discussions of this success of Koide's formula (\ref{KoideFormula}) are naturally 
formulated in a $Z_3$-symmetric framework
by parametrizing the masses of any three given fermions $f_1,f_2,f_3$
in terms of three parameters $M_f,k_f,\delta_f$
 as  \cite{Koide2,Brannen}:
\begin{eqnarray}
\label{KoideParametrization}
\sqrt{m_{f_j}}&=&\sqrt{M_f}\,\left(1+\sqrt{2}\,k_f 
\cos {\left(\frac{2 \pi j}{3}+\delta_f\right)}\right), ~~~~~~~~~(j=1,2,3).
\end{eqnarray} 
This  choice of parametrization of masses
is particularly suited to Koide's formula 
as not only $M_f$ but also
$\delta_f$ drop out of the r.h.s. of Eq. (\ref{KoideFormula}). \\

Since $\delta_f$ is free we may 
assume $m_1\le m_2 \le m_3$ without any loss of generality.
From Eq. (\ref{KoideParametrization}) one then gets a counterpart of
Eq. (\ref{KoideFormula}), in which it is now $k_L$ that drops out of the 
formula:
 \begin{equation}
 \label{tandeltaL}
 \frac{\sqrt{3}\,(\sqrt{m_{\mu}}-\sqrt{m_e})}
{2\sqrt{m_{\tau}}-\sqrt{m_{\mu}}-\sqrt{m_e}}=\tan \delta_L.
 \end{equation}
From the experimental values 
of $e$, $\mu$ and $\tau$ masses one finds:
\begin{equation}
\label{deltaLexp}
\delta_L=0.2222324,
\end{equation}
which, as observed by Brannen and Rosen \cite{Brannen2,Rosen},
 is extremely close to 
 \begin{equation}
 \label{deltaL29}
 \delta_L=2/9.
 \end{equation}
Conversely, assuming $\delta_L=2/9$, Eq. (\ref{tandeltaL}) predicts 
the value of the tauon mass in terms of experimental $e$ and $\mu$ 
masses. Just as in the case of Koide's formula, the relevant prediction is 
within one standard deviation from the measured $\tau$ mass:
\begin{equation}
\label{tauRosenprediction}
m_{\tau}({\rm \delta_L=2/9})=1776.9664 ~{\rm MeV}.
\end{equation}
Assuming that the Koide and the Brannen-Rosen observations do not reflect mere
coincidences, the $Z_3$-symmetric parametrization (\ref{KoideParametrization})
should be rightly called `doubly special'. A peculiar feature of this
parametrization is that the simple numbers of $1$ and $2/9$ work well at the 
{\it low}-energy scale and not at some high mass scale. For example, taking the values
of charged-lepton masses at the mass scale of
$M_Z$, the extracted values of $k_L$ and $\delta_L$ deviate from their
`perfect' values of $1$ and $2/9$ by about $0.2$ \% and $0.5$ \% respectively.
Apparently, an explanation of the success of predictions
(\ref{tauKoideprediction}, \ref{tauRosenprediction}) should not be sought
 at the high-mass scale of some grand unified theory (see eg. \cite{Mohapatra}).

\section{Extending the scheme to the quark sector}
If there is some physical reason behind the appearance of simple numbers
such as $1$ and $2/9$ 
in the charged-lepton sector, one would expect its analogs 
working in the quark and neutrino sectors as well.
However, it is known that the original Koide formula (\ref{KoideFormula}) 
does not work when 
replacing the charged-lepton masses with those of  
neutrinos or quarks. For neutrinos one estimates directly
from experiment that $k_{\nu} \leq 0.81 $ \cite{RodejohannZhang} (the
mathematically allowed region being $0 \leq k_f \leq \sqrt{2}$).
For quarks, using their mass values appropriate at $\mu = 2~GeV$,
one obtains $k_D \approx 1.08$ ($k_U\approx 1.25$) for the down (up) quarks 
respectively
\cite{RodejohannZhang,XingZhang}. If a higher energy scale
 $\mu=M_Z$ is taken, even larger values are obtained, i.e. 
 $k_D=1.12$ and $k_U=1.29$.
Going from $\mu=2~GeV$ towards the low energy scale leads to smaller values 
of $k_D$ and $k_U$. However, the top quark mass is so large that one certainly
cannot bring $k_U$ into the vicinity of $1$. \\

On the other hand, it has been observed recently \cite{ZenKoide1} from 
the quark sector analogs
of Eq. (\ref{tandeltaL}) that at the {\it low}-energy scale 
the relevant phase parameters acquire approximate values:
\begin{eqnarray}
\delta_U& \approx & 2/27=\delta_L/3, \nonumber \\
\label{deltaUD}
\delta_D& \approx & 4/27=2\delta_L/3.
\end{eqnarray}
Due to the problem of quark confinement we obviously cannot
 check how precise the above equalities are. However, 
 given the accuracy of their lepton counterpart 
  (Eq. (\ref{tauRosenprediction})) we may {expect} that they are nearly 
 {exact}. Therefore, we will assume this from now on and {\it conjecture} that
at the low energy scale the charged-lepton and quark mass bases are 
 characterised by Eqs. (\ref{deltaL29}, \ref{deltaUD}).

 The problem then remains how to interpret the value of $k_L=1$ and how it should
 be generalized to the quark sector.
 Here we accept the suggestion of ref. \cite{Gerard} that $k_f=1$ is a feature
 of the weak basis. Thus,
 according to \cite{Gerard}, the masses of charged leptons are described by 
 $k_L=1$ because for charged leptons 
 the mass and the weak bases coincide. The
 lepton mixing matrix (for simplicity we assume here Dirac neutrino masses),
 i.e.
 \begin{equation}
 V_{MNS}=U_L^{\dagger}U_{\nu},
 \end{equation}
 is then wholly assigned to the contribution from neutrinos (for which $k_{\nu}\ne
 1$):
 \begin{eqnarray}
 U_{\nu}&=&V_{MNS},\nonumber \\
 \label{leptondecomposition}
 U_L&=&1.
 \end{eqnarray}
 In other words, if $U_L$ were different from $1$, one would not expect
 the simplicity of Koide's formula to persist.\\
 
 Since the analogs of the charged-lepton equality $k_L=1$ do not hold 
 in the $U$ and $D$ quark sectors, one expects that the  quark counterpart 
 of the  decomposition
 (\ref{leptondecomposition}) will be also modified.
 Thus,  in the CKM matrix 
 \begin{equation}
 V_{CKM}=U_U^{\dagger}U_{D},
 \end{equation}
  both of the factor matrices $U_U$ and $U_D$ are expected to be different 
  from $1$.
  
  \noindent
  For fermion type $f$ the general connection between the mass and the weak bases is
 \begin{equation}
 {\rm diag}(m_{f_1},m_{f_2},m_{f_3})=U^{\dagger}_{f,left}MU_{f,right},
 \end{equation}
 where $M$ is the mass matrix in the weak basis. In the above formula
  $U_{f,left}\equiv U_f$,  
 while $U_{f,right}$ may be chosen equal to $1$.
 As a result, in the weak basis one deals with `pseudo-masses' 
 $\tilde{m}_{f_j}$ defined as \cite{Gerard}
 \begin{equation}
 \label{pseudomasses}
 \tilde{m}_{f_j}=|\sum_lU_f^{jk}m_{f_k}|.
 \end{equation}
 For charged leptons ($f=L$, $U_L=1$) these pseudo-masses coincide with the 
 observed  masses, i.e. $\tilde{m}_{L_j}=|m_{L_j}|$. On the other hand, for 
 quarks  ($f=D,U$ with $U_{D}, U_U \ne 1$) the pseudo-masses are different from 
 the  observed mass values.
 It is for these `pseudo-masses' that, 
 according to the proposal of \cite{Gerard}, the analogs of Koide's formula 
 (\ref{KoideFormula}) are supposed to hold with $k_D=k_U=1$.
 The authors of ref. \cite{Gerard} used quark masses at the $Z$ mass scale
 and found that it is possible to get
 $k_D=k_U \approx 1$ provided one takes a value
 of the strange quark mass $m_s(Z)$ that is larger by a factor of 2.5 (!)
 from the theoretical estimate at that scale.
 Since the Koide and the Brannen-Rosen observations work best 
 at the {\it low}-energy scale (and not at the $Z$ mass),
 and since at the low-energy scale the masses of quarks (and especially $m_s$, 
 see \cite{ZenKoide1}) are naturally expected to be
 larger than at the $Z$ mass, 
 a question appears if it is possible to recover Koide's formula for quarks
 using low-energy quark masses corresponding to phases of formulas (\ref{deltaUD}).
 This is the question asked here. Thus, the present paper constitutes a
 low-energy-scale study of the idea of ref.\cite{Gerard}.

 \section{The structure of $U_f$ in the quark sector}
  As the assumption of Koide's formula for pseudo-masses 
  imposes constraints upon matrices   $U_{D}$ and $U_U$ (and,
 consequently, upon  $V_{CKM}$), we have to discuss these matrices
 in some detail.

 In \cite{FritzschXing1} Fritzsch and Xing convincingly argued that the hierarchical structure of quark
 mass terms suggests certain particular parametrizations of $U_D$ and $U_U$ 
 as `most physical' (i.e. that it selects
 one of the nine possible parametrizations of the $V_{CKM}$ 
 matrix  \cite{FritzschXing2} as probably the
 most suitable for the description of the quark-mixing phenomenon). 
 The same parametrization was also advocated in ref. \cite{Gerard} where the
 conjecture that Koide's formula holds in the weak basis was originally formulated.
 Consequently, we think that it is justified to accept the Fritzsch-Xing
 parametrization here.
 The relevant `natural' parametrizations of $U_D$ and $U_U$ are then:
 \begin{eqnarray}
U_D&=&R_{23}(\phi_b,\theta_b)\,R_{12}(\theta_d),\nonumber \\
\label{UdUuparametrizations}
U_U&=&R_{23}(\phi_t,\theta_t)\,R_{12}(\theta_u),
\end{eqnarray}
with ($c_q \equiv \cos \theta_q$, $s_q \equiv \sin \theta_q$)
\begin{eqnarray}
R_{12}(\theta_{q})~=&~
\left( 
\begin{array}{ccc}
c_{q} & -s_{q} & 0\\
s_{q} & c_{q} & 0\\
0 & 0 & 1
\end{array}
\right),~~~~~~~~~~~~ & ~~~~(q=d,u),\\
\label{R23form}
R_{23}(\phi_q,\theta_q)~=&
\left(
\begin{array}{ccc}
e^{-i\phi_q} & 0 & 0\\
0& c_q & s_q \\
0& -s_q& c_q 
\end{array}
\right), ~~~~~~& ~~~~(q=b,t).
\end{eqnarray}
Thus, the induced parametrization of the $V_{CKM}$ matrix can be read from:
\begin{equation}
V_{CKM}=R_{12}^{\dagger}(\theta_u)\,R_{23}^{\dagger}(\phi_t,\theta_t)\,
R_{23}(\phi_b,\theta_b)\,R_{12}(\theta_d).
\end{equation}
The product $R_{23}^{\dagger}(\phi_t,\theta_t)R_{23}(\phi_b,\theta_b)$
may be written in the form of Eq. (\ref{R23form}) 
with a single phase
$\phi=\phi_b-\phi_t$
and a single rotation angle $\theta=\theta_b-\theta_t$, 
(with $c \equiv \cos \theta$, $s \equiv \sin \theta$)
\begin{equation}
R_{23}^{\dagger}(\phi_t,\theta_t)R_{23}(\phi_b,\theta_b)\equiv 
R_{23}(\phi,\theta)=\left(
\begin{array}{ccc}
e^{-i\phi} & 0 & 0\\
0& c & s \\
0& -s & c 
\end{array}
\right).
\end{equation}
The CKM matrix is then parametrized as
\begin{eqnarray}
V_{CKM}&=&
\left( 
\begin{array}{ccc}
c_{u} & s_{u} & 0\\
-s_{u} & c_{u} & 0\\
0 & 0 & 1
\end{array}
\right)
\left(
\begin{array}{ccc}
e^{-i\phi} & 0 & 0\\
0& c & s \\
0& -s & c 
\end{array}
\right)
\left( 
\begin{array}{ccc}
c_{d} & -s_{d} & 0\\
s_{d} & c_{d} & 0\\
0 & 0 & 1
\end{array}
\right)\nonumber \\
\label{CKMparametrization}
&=&
\left(
\begin{array}{ccc}
s_us_dc+c_uc_de^{-i\phi}~~&s_uc_dc-c_us_de^{-i\phi}~~&s_us\\
c_us_dc-s_uc_de^{-i\phi}~~&c_uc_dc+s_us_de^{-i\phi}~~&c_us\\
-\, s_ds~~&- \,c_ds~~&c
\end{array}
\right).
\end{eqnarray}
Since quark fields can be freely rephased, all of the
three angles $\theta_d, \theta_u, \theta$ can be arranged to lie in the
first quadrant so that $s_u, s_d, s$ and $c_u, c_d, c$ are all positive
(the phase $\phi$ cannot be so restricted). 
 
 We use the following absolute values of the elements of 
 the CKM matrix  most relevant for our parametrization \cite{PDG}:
 \begin{eqnarray}
 |V_{ub}|&=&0.00351\pm0.00015,\nonumber \\
 |V_{cb}|&=&0.0412\pm0.0008,\nonumber \\
 |V_{td}|&=&0.00867\pm0.00030,\nonumber \\
 |V_{ts}|&=&0.0404\pm0.0008,\nonumber\\
 |V_{tb}|&=&0.999146\pm0.000034.
 \end{eqnarray}
Inserting the above numbers, we find from (\ref{CKMparametrization}):
\begin{eqnarray}
\theta_u&=&4.87^{\circ}\pm0.23^{\circ},\nonumber \\
\theta_d&=&12.11^{\circ}\pm0.47^{\circ},\nonumber \\
\label{anglesexp}
\theta=\theta_b-\theta_t&=&2.37^{\circ}\pm0.05^{\circ}.
\end{eqnarray} 
Since formula (\ref{pseudomasses}) involves absolute values, the actual sizes
 of $\phi_b$ and $\phi_t$ and, consequently, the experimentally imposed 
restriction on the CP-violating phase parameter $\phi=\phi_b-\phi_t$
 are irrelevant for our purposes.

\section{Imposing Koide's condition on pseudo-masses}
The values of $\delta_D=4/27$ and $\delta_U=2/27$, together with the low-energy
ratios $m_s/m_d=20.4$, $m_u/m_d=0.56$ (see e.g. \cite{Weinberg}) 
suffice to fix the 
pattern of low-energy quark masses
up to two overall mass scales (in the up and down sectors). 
These scales are irrelevant for the discussion 
of Koide's formulas for pseudo-masses. For illustrative purposes, however,
one may set $m_s=160.0~MeV$ and $m_t=172000~MeV$.
This choice leads to the following representative values of 
{\it low}-energy quark masses (in $MeV$):
\begin{eqnarray}
m_d=7.843,~~&m_s=160.0,~~&m_b=4209,\nonumber \\
m_u=4.392,~~&m_c=1296,~~&m_t=172000.
\end{eqnarray}
\begin{figure}[t]
\caption{Correlations required by Eq. (\ref{Koidepseudo}):
(a) $\theta_b \leftrightarrow \theta_d$ and 
 (b) $\theta_t \leftrightarrow \theta_u$ 
(all angles in degrees).  
Solid lines correspond to $k_D=k_U=1$. 
Dashed lines denote solutions $\theta_{b,1}$ and $\theta_{t,1}$ for 
$k_D=k_U=1.015$. 
}
\label{fig1}
\begin{center}
\epsfxsize=7.2 cm
\mbox{\epsfbox{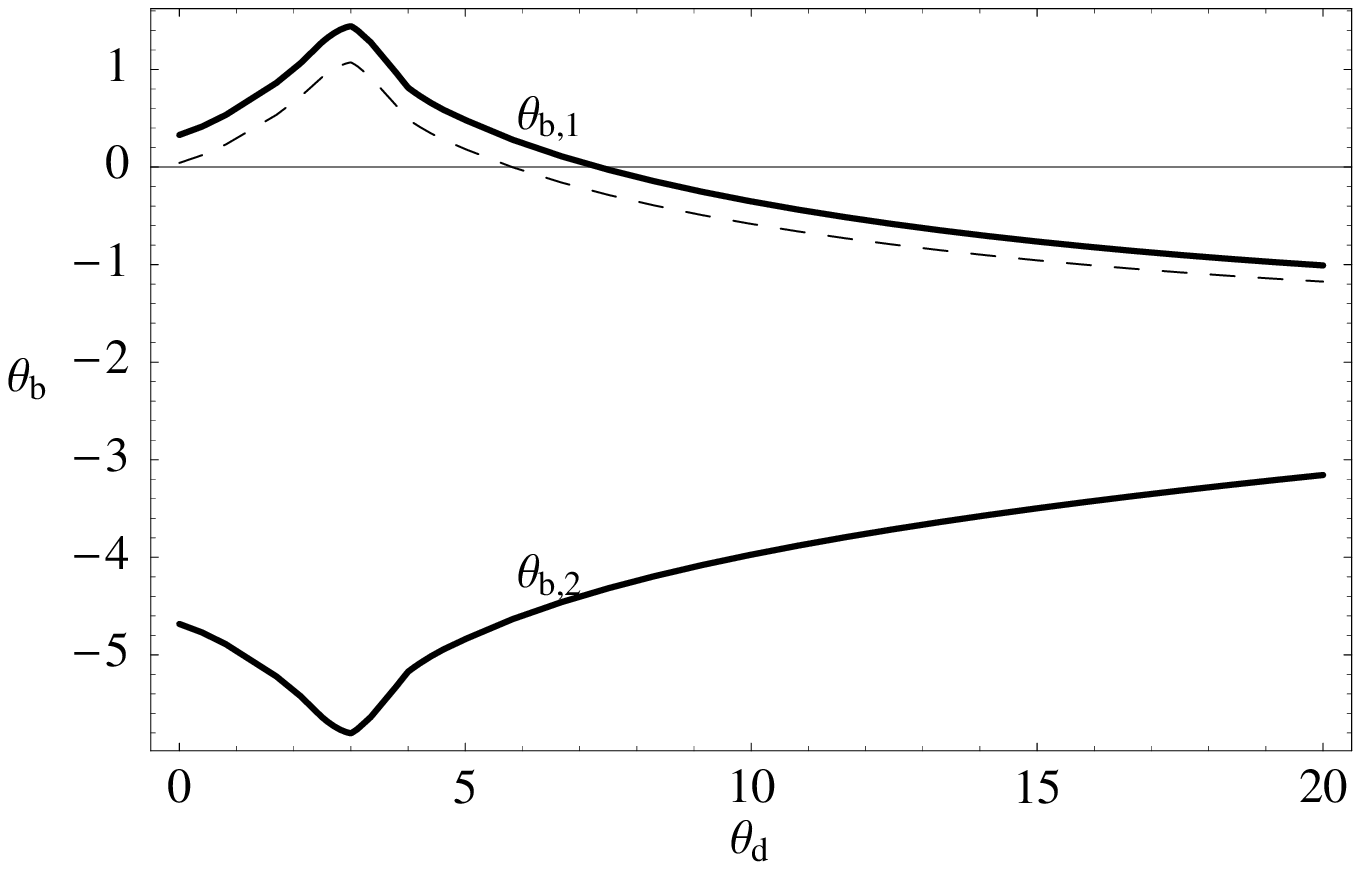}} 
\epsfxsize=7.2 cm 
\mbox{\epsfbox{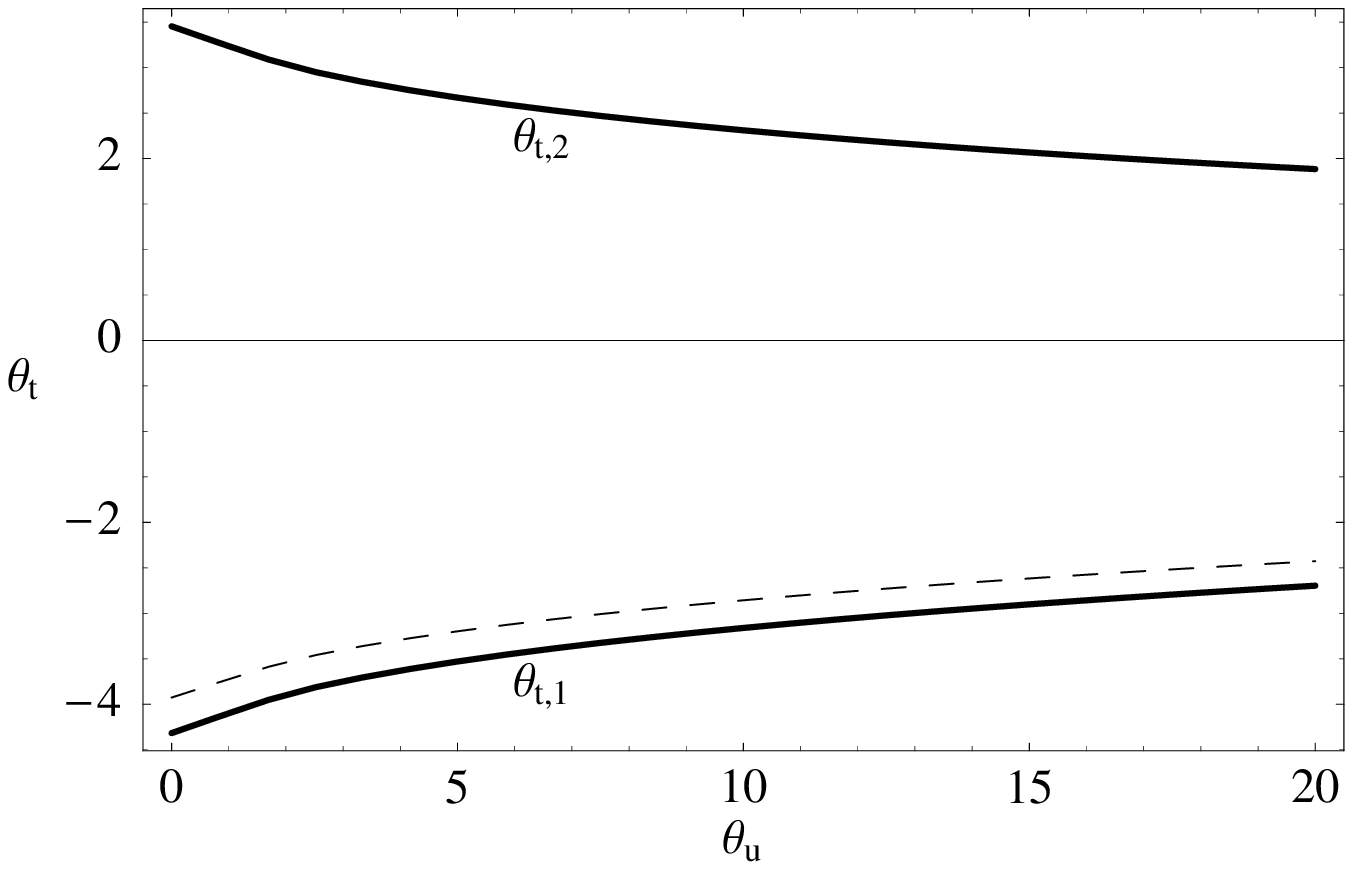}}\\
~~~~~(a)~~~~~~~~~~~~~~~~~~~~~~~~~~~~~~~~~~~~~~~~~~~~~~~~~(b)
\end{center}
 \end{figure}

\begin{figure}[t]
\caption{Contour plots of 
$\theta=\theta_{b,1}(\theta_d)-\theta_{t,1}(\theta_u)$ (all angles in degrees): 
(a) $k_D=k_U=1$ and (b) $k_D=k_U=1.015$ . 
Dashed lines are contours corresponding to 
$\theta=2.37^{\circ}\pm 0.10^{\circ}$. 
Cross denotes
experimental point $(\theta_d,\theta_u)=(12.11^{\circ}\pm 0.47^{\circ},
4.87^{\circ}\pm 0.23^{\circ})$. 
}
\label{fig2}
\begin{center}
\epsfxsize=7.2 cm
\mbox{\epsfbox{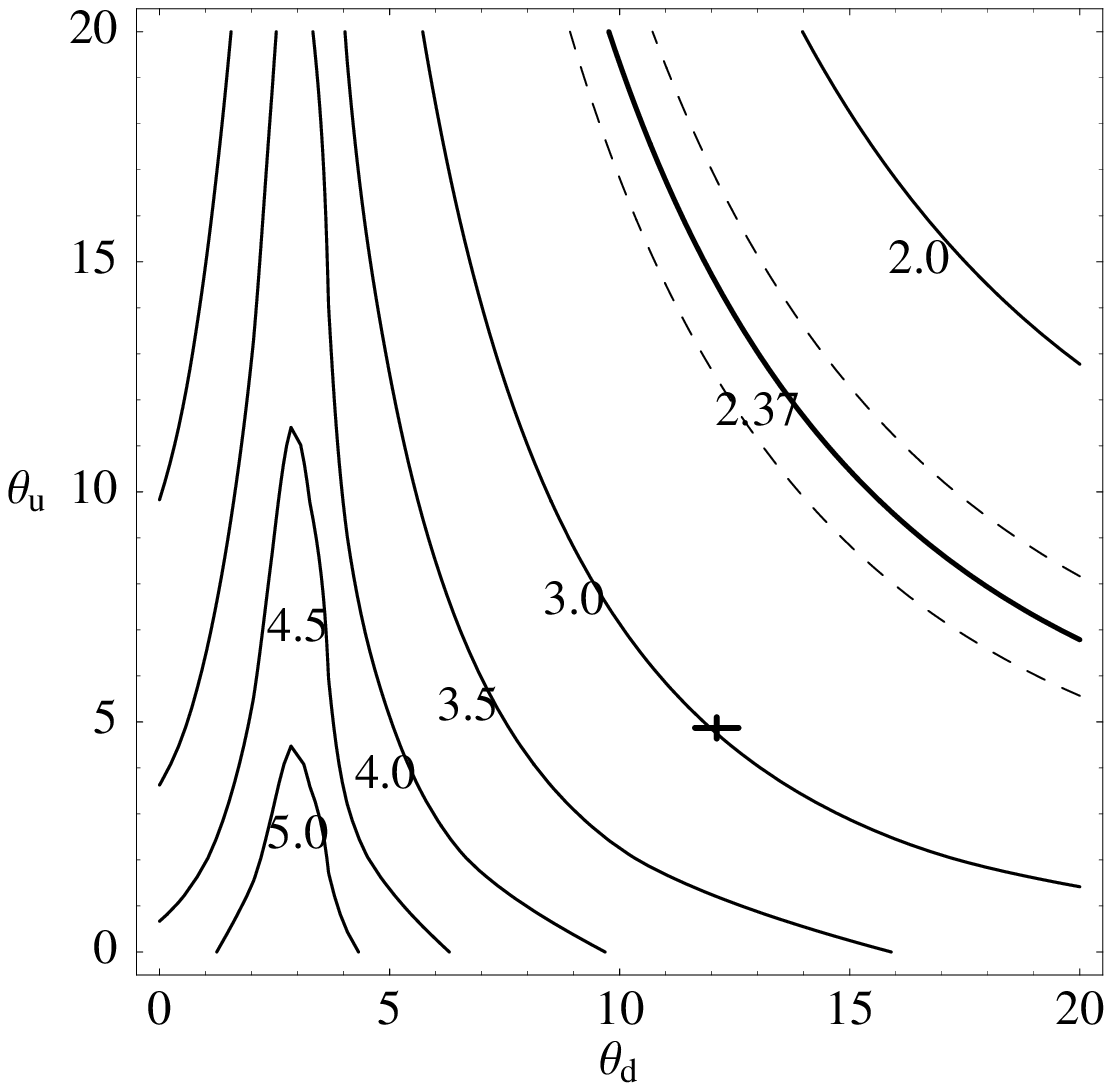}} 
\epsfxsize=7.2 cm 
\mbox{\epsfbox{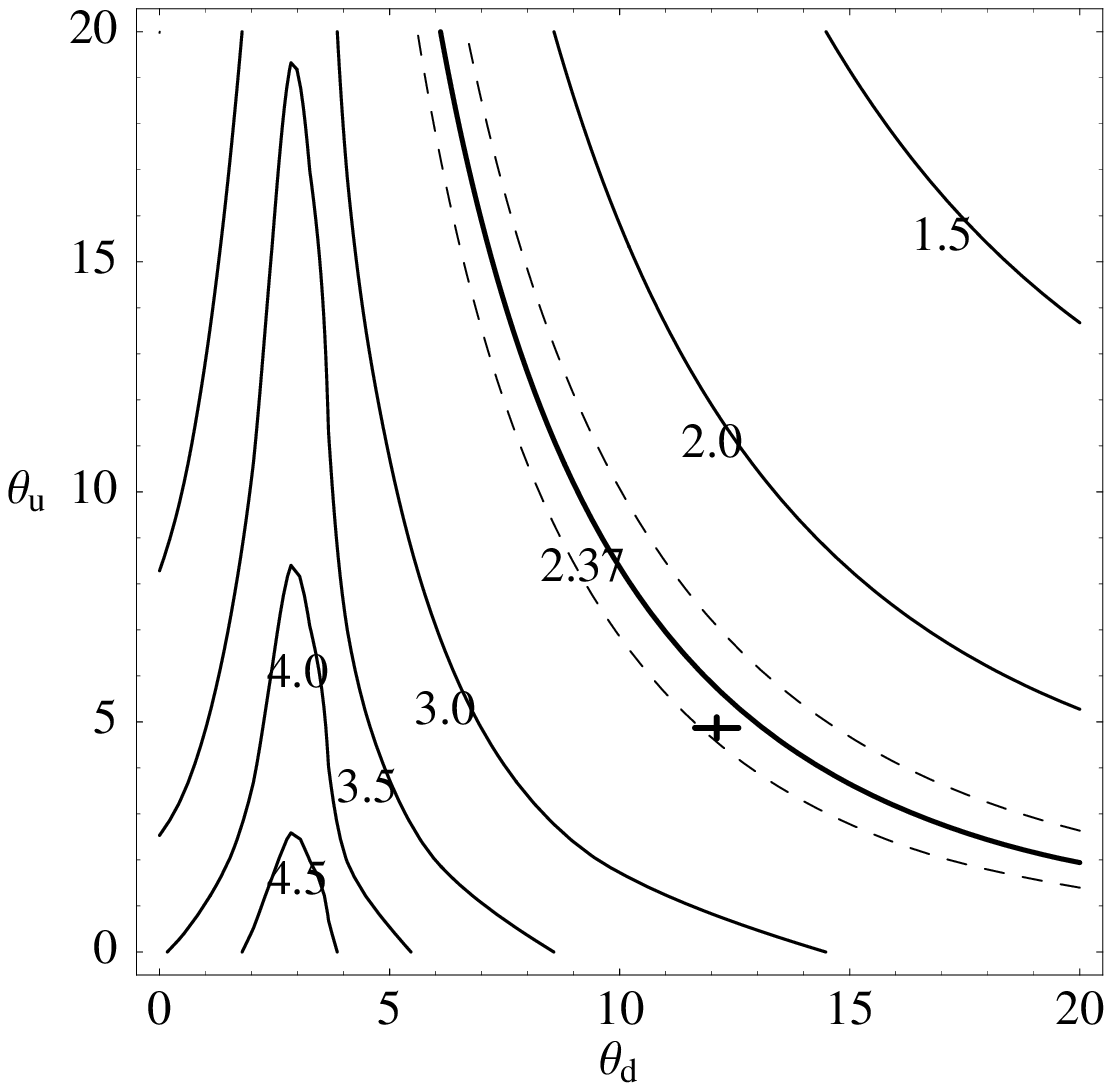}}\\
~~~~~(a)~~~~~~~~~~~~~~~~~~~~~~~~~~~~~~~~~~~~~~~~~~~~~~~~~(b)
\end{center}
 \end{figure}
 For a given value of $k_f$,
 with $\phi_b$, $\phi_t$ dropping from expression (\ref{pseudomasses}) 
for the pseudo-masses,  the condition 
\begin{equation}
\label{Koidepseudo}
\frac{\sum_j \tilde{m}_{f_j}}
{\left(\sum_j \sqrt{\tilde{m}_{f_j}}\right)^2}=\frac{1+k_f^2}{3}
\end{equation}
imposes two constraints: between $\theta_d$ and $\theta_b$ in the down quark sector,
and between $\theta_u$ and $\theta_t$ in the up quark sector.
Thus, $\theta_b$  becomes dependent on 
$\theta_d$ (and $\theta_t$ on $\theta_u$). 
For $k_D=k_U=1$  there are two possible solutions
for function $\theta_b(\theta_d)$ and two possible solutions for 
function $\theta_t(\theta_u)$. 
They are shown in Fig. \ref{fig1} with solid lines 
 marked as
$\theta_{b,n}$ and $\theta_{t,n}$ ($n=1,2$).
However, only the combination $\theta_{\rm Koide}=\theta_{b,1}-\theta_{t,1}$ 
is positive. Specifically,
taking the central experimental values of $\theta_u$ and $\theta_d$,  
one finds
\begin{equation}
\theta_{\rm Koide}=2.98^{\circ},
\end{equation}
which is only slightly different from the value given in 
Eq. (\ref{anglesexp}). 

A question thus emerges 
how big a departure of $k_D,k_U$ from 1 is needed
to fit 
the current experimental value of $\theta$.
 The dashed lines in Fig. \ref{fig1} correspond to $\theta_{b,1}$ and 
 $\theta_{t,1}$
as obtained for $k_D=k_U=1.015$. The predicted value of $\theta$ is then
$\theta(k_D=k_U=1.015) \approx 2.44^{\circ}$, 
which is in good agreement with experiment.
For completeness, 
we also show the contour plot of $\theta$ as a function of $\theta_d$
and $\theta_u$ for $k_D=k_U=1$ (Fig. \ref{fig2}a) and $k_D=k_U=1.015$ 
(Fig. \ref{fig2}b). Dashed lines
correspond to contours $2.37^{\circ} \pm 2 \sigma $ (with
$\sigma=0.05^{\circ}$).

Although it might seem that the above results indicate that
one cannot obtain $k_U=k_D=1$, this is not the case.
One has to remember that while parametrizations (\ref{UdUuparametrizations})
have been suggested as the most appropriate ones \cite{FritzschXing1}, 
they may be `naturally' modified. 
Indeed, the same $V_{CKM}$ is obtained if one substitutes
$U_D\to U'_D=W_DU_D$ and $U_U\to U'_U=W_UU_U$, provided $W_D$ and $W_U$ denote
the same
arbitrary unitary matrix. 
If $W_D\ne W_U$, but both $W_D$ and $W_U$ are very close to $1$, 
a  minor modification of our results is expected. Such a natural modification 
of $U_{D(U)}=R^{D(U)}_{23}R^{D(U)}_{12}$ is
 obtained if we put $W_{D(U)}=R^{D(U)}_{13}(\theta^{D(U)}_{13})$ 
 with very small $\theta^D_{13}\ne \theta^U_{13}$. Since the inclusion of 
 two new parameters $\theta^{D(U)}_{13}$ introduces additional freedom
 into the scheme, it does not make sense to study it here. However -
 keeping this freedom in mind -
the parametrization of Eq.(\ref{UdUuparametrizations}) and our numerical
results may be viewed as capturing the dominant effects only.

In conclusion, the data are consistent with the statement
 that low-energy quark masses satisfy phase relations $\delta_D=2\delta_U=4/27$, 
 while the expected Koide relations $k_U=k_D = 1$ 
hold approximately for masses transformed to the weak basis, as suggested in
\cite{Gerard}.
These observations might be relevant for a future theory of mass.

\vfill

\vfill

\end{document}